\documentstyle[preprint,epsfig,aps,prl]{revtex}

\tightenlines
\title{Barrier-controlled carrier transport in microcrystalline \\
semiconducting materials: Description within a unified model}
\author{T.\ Weis $^{\rm a}$, R.\ Lipperheide $^{\rm
a}$, U.\ Wille $^{\rm a,}$\footnote{Author to whom correspondence should be
addressed; electronic mail: wille@hmi.de}, and  S.\ Brehme $^{{\rm b}}$ \\[0.5cm]}
\address{$^{\rm a}$ Abteilung Theoretische Physik, Hahn-Meitner-Institut
Berlin, Glienicker Str.\ 100,\\ D-14109 Berlin, Germany\\$^{\rm b}$
Abteilung Silizium-Photovoltaik, Hahn-Meitner-Institut Berlin, Kekul\'{e}str.\
5,\\ D-12489 Berlin, Germany}

\begin{document}
\date{\today}
\maketitle

\begin{abstract}
A recently developed model that unifies the ballistic and diffusive transport
mechanisms is applied in a theoretical study of carrier transport across
potential barriers at grain boundaries in microcrystalline semiconducting
materials.  In the unified model, the conductance depends on the detailed
structure of the band edge profile and in a nonlinear way on the carrier mean
free path.  Equilibrium band edge profiles are calculated within the trapping
model for samples made up of a linear chain of identical grains.  Quantum
corrections allowing for tunneling are included in the calculation of electron
mobilities.  The dependence of the mobilities on carrier mean free path, grain
length, number of grains, and temperature is examined, and appreciable
departures from the results of the thermionic-field-emission model are found.
Specifically, the unified model is applied in an analysis of Hall mobility data
for n-type $\mu$c-Si thin films in the range of thermally activated transport.
Owing mainly to the effect of tunneling, potential barrier heights derived from
the data are substantially larger than the activation energies of the Hall
mobilities.  The specific features of the unified model, however, cannot be
resolved within the rather large uncertainties of the analysis.
\\[0.3cm] PACS
Numbers:  72.20.-i, 73.50.-h, 73.50.Dn, 73.50.Jt

\end{abstract}

\newpage

\section{Introduction}
Over the past decades, a great deal of interest has been devoted to the study
of the transport properties of microcrystalline semiconducting materials
\cite{ort80a,gro85,kam86,wil91,ica00}.  The commonly adopted viewpoint is that
these properties are largely determined by potential barriers built up at grain
boundaries due to charge carrier trapping into interface states \cite{tay52}.
Carrier transport across these barriers is mostly described in terms of the
traditional thermionic-emission model \cite{bet42}, or the
thermionic-field-emission model \cite{dol54,str62} in which corrections
allowing for quantum tunneling through the barriers are included.

Systematic studies of the transport properties in microcrystalline materials
are usually based on measurements of Hall mobilities in thin films
\cite{ort80a,wil91}.  Seto \cite{set75} has been the first to perform a
detailed theoretical analysis of Hall data, applying the grain-boundary
trapping model in conjunction with the thermionic-emission model.  For p-type
$\mu$c-Si thin films, potential barrier heights and trapping-state densities
were deduced for different doping concentrations.  In a large number of
subsequent studies (see, e.g., Refs.\
\cite{ort80,rot81,ric81,nak93,dhe94,pri98,nak99,car99,bre01a,bre01b}, and
references cited therein), Hall mobilities were analyzed with regard to their
dependence on temperature, doping concentration, and film thickness.  While, in
general, these analyses have supported the picture of thermally activated
transport over grain boundary barriers, the quantitative understanding of
barrier-controlled transport appears to be incomplete yet.  For example, the
effect of tunneling \cite{rot81} and the detailed form of the density of
grain-boundary trapping states \cite{pik79,wer89} need further clarification.

In the present paper, we concentrate on one particular aspect of the decription
of barrier-controlled transport that does not seem to have been systematically
studied until now, viz., the role played by the transport mechanism.  In many
practical cases, the magnitude of the carrier mean free path in
microcrystalline material is comparable to the grain length, and so it appears
indicated to consider departures of the transport mechanism from the commonly
assumed thermionic-emission mechanism.  A framework well suited to deal with
this situation is provided by the unified transport model developed recently by
the present authors \cite{lip01a}.  This model unifies the ballistic and
diffusive transport mechanisms in a generalization of the Drude model.  The
unified model is valid for arbitrary shape of the band edge profiles and for
arbitrary magnitude of the carrier mean free path.

In the next section, the main features of the unified transport model are
briefly summarized and discussed.  In Sec.\ III, electron mobilities calculated
within the unified model are presented and analyzed with regard to their
dependence on various parameters.  Comparison is made with the
thermionic-field-emission model.  Section IV deals with the application of the
unified model in an analysis of Hall mobilities for n-type $\mu$c-Si
thin films.  By suitably choosing the model parameters, grain-boundary
potential barrier heights are inferred from the Hall data.  Finally, in Sec.\
V, the contents of the paper are summarized and some concluding remarks are
made.

\section{Unified transport model}
In this section, we summarize and discuss the essential features of the unified
transport model \cite{lip01a}.  A one-dimensional formulation is used in
conjunction with the semiclassical approach.  We restrict ourselves to
considering the case of electrons (holes are described analogously).

The unified model is based on the idea that electrons move ballistically in the
electric field over intervals with average length equal to a universal mean
free path $l$.  At the end of a ballistic interval, the electrons are
thermalized into a state of local equilibrium characterized by a quasi-Fermi
level $E_{\rm F}(x)$.  The length $S$ of the sample is made up of random
configurations of ballistic intervals.  By appropriately averaging over these
configurations, one arrives at a unified description of electron transport,
which is valid for arbitrary magnitude of $l$ and for arbitrary shape of the
conduction band edge profile $E_{\rm c}(x)$.  The purely ballistic and purely
diffusive transport mechanisms appear as limiting cases of this description.

For the present purpose, the principal result of the unified model is the
formula for the zero-bias conductance per unit area,
\begin{equation}
g = \frac{4\pi e^{2} m^{*}}{\beta h^{3}} \, \ln (1+ e^{-\beta E_{\rm p}})
\, \Gamma \, \frac{l}{L}
\label{eq:1}
\end{equation}
$(\beta = 1/k_{\rm B} T)$, which allows for degeneracy and quantum tunneling.
This formula comprises Eqs.\ (46) and (48) of Ref.\ \cite{lip01a}, in which
degeneracy and tunneling corrections, respectively, were taken into account
separately.  In Eq.\ (\ref{eq:1}), $E_{\rm p} = E_{\rm c}^{\rm m} - E_{\rm F}$,
where $E_{\rm c}^{\rm m}$ is the maximum of the band edge profile $E_{\rm
c}(x)$ for $0 \leq x \leq S$ and $E_{\rm F}$ is the Fermi level for zero bias
(note that in the degenerate case, $E_{\rm p}$ may be negative).  The quantity
$\Gamma$ is the tunneling correction, and $L$ is the ``effective transport
length''. In accordance with our description of carrier transport along
classical trajectories, we treat tunneling in WKB approximation.

\begin{figure}
\vspace*{2.1cm}
\epsfysize=8.0cm
\epsfbox[40 671 257 782]{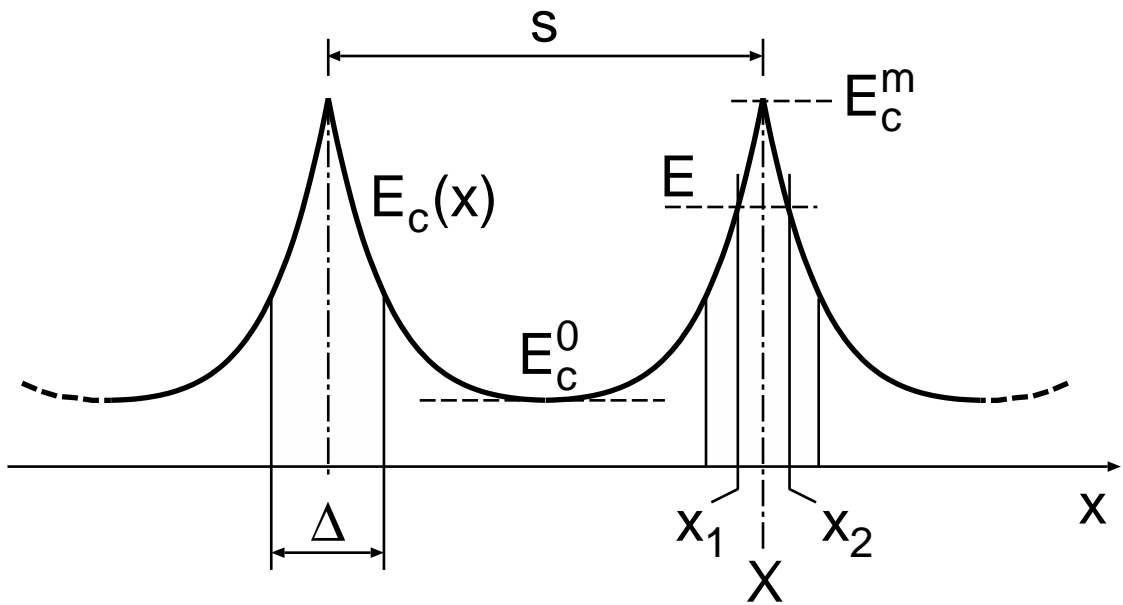}
\vspace*{1.7cm}
\caption{Schematic equilibrium band edge profile for a chain of identical
grains.}
\label{fig:1}
\end{figure}

As we deal here with chains of {\em identical} grains, it is sufficient to
consider tunneling through a single, symmetric potential barrier with maximum
$E_{\rm c}^{\rm m}$ at $x = X$ (cf.\ Fig.\ \ref{fig:1}).  The correction
$\Gamma$, defined as the ratio of the energy-integrated WKB and classical
barrier transmission probabilities, is then given by
\begin{equation}
\Gamma = 1 +  [\ln (1+ e^{-\beta E_{\rm p}})]^{-1} \, \beta \int_{E_{\rm
c}^{0}}^{E_{\rm c}^{\rm m}} dE \, \frac{T^{\rm WKB}(E)}{1 + e^{\beta [E
- E_{\rm F}]}} \; ,
\label{eq:2}
\end{equation}
where $E_{\rm c}^{0} = E_{\rm c}(X - s/2)$ is the value of the band edge
profile at the center of the grain. For the WKB tunneling probability $T^{\rm
WKB}(E)$, we have
\begin{equation}
T^{\rm WKB}(E) = \exp\left( -(2/\hbar) \int_{x_{1}}^{x_{2}} dx \{2m^{*}[E_{\rm
c}(x) - E] \}^{1/2} \right) .
\label{eq:3}
\end{equation}
Here, $x_{1}$ and $x_{2}$ are the positions of the left and right turning
point, respectively, of the classical electron motion at total energy $E$.

The effective transport length $L$ has the form
\begin{equation}
L = l + \widetilde{S} + \widetilde{\Lambda} \; .        \label{eq:4}
\end{equation}
It comprises, aside from the mean free path $l$ which represents the ballistic
contribution to electron transport, the ``reduced sample length''
$\widetilde{S}$ and the ``shape term'' $\widetilde{\Lambda}$.  While
$\widetilde{S}$ reflects the diffusive contribution, $\widetilde{\Lambda}$
represents the interplay between ballistic and diffusive transport.  For a
chain of $\nu$ identical grains, each of length $s$, the reduced sample
length is given by
\begin{equation}
\widetilde{S} =  \nu \; \widetilde{s} = \nu \, \Gamma \int_{X - s}^{X} dx \,
F(x) \; , \label{eq:5}
\end{equation}
where
\begin{equation}
F(x) =  \frac{\ln (1+ e^{\beta [E_{\rm F} -
E_{\rm c}^{\rm m }]})}{\ln (1+ e^{\beta [E_{\rm F} - E_{\rm c}(x)]})} \; .
\label{eq:6}
\end{equation}
For the shape term, we have
\begin{eqnarray}
\widetilde{\Lambda} = (\nu -1) \; \widetilde{\lambda} &=&  (\nu -1) \, \int_{X
-s}^{X} dx \, e^{-2|x-(X-s/2)|/l} \left[ 1 - \Gamma \, F(x) \right] \nonumber
\\[0.3cm] &=&(\nu -1) \left\{ (1-e^{-s/l})\, l - \Gamma \,\int_{X-s}^{X} dx \,
e^{-2|x-(X-s/2)|/l} \, F(x) \right\} \; .
\label{eq:7}
\end{eqnarray}

It is seen from Eq.\ (\ref{eq:1}) that, except for an overall factor, the
quantity $R = \Gamma^{-1}\,  (L/l)$ is the resistance of the chain. We split up
$R$ into a net ballistic contribution $R_{\rm b}$ and a net diffusive
contribution $R_{\rm d}$,
\begin{equation}
R = \Gamma^{-1} \, \frac{L}{l} = R_{\rm b} + R_{\rm d} \; .
\label{eq:8}
\end{equation}
Using Eqs.\ (\ref{eq:4}), (\ref{eq:5}), and (\ref{eq:7}), we find for the
{\em ballistic} contribution  (which is proportional to the inverse of the
probability for transmission through the barriers)
\begin{equation}
R_{\rm b} = \Gamma^{-1} \, [1 + (\nu -1) (1 - e^{-s/l}) ] \; .
\label{eq:9}
\end{equation}
In this expression, one barrier contributes fully ballistically, while the
contribution of the other $\nu - 1$ barriers is weighted by the probability $1
- \exp(-s/l)$ that one or more collisions (thermalizations) occur within the
distance $s$ in the valley between one barrier and the next.  Without such
collisions, the barrier at the end of the valley would not contribute; it would
be ``eclipsed'' by the preceding barrier (cf.\ Ref.\ \cite{lip01a}).  The
contribution of each barrier contains the factor $\Gamma^{-1}$, which takes
into account the reduction of the resistance arising from tunneling. For the
{\em diffusive} contribution, we have
\begin{equation}
R_{\rm d}  = \int_{X-s}^{X} \frac{dx}{l} \, [\nu - (\nu -1) e^{-2|x-(X
-s/2)|/l}] \, F(x) \; .
\label{eq:10}
\end{equation}
Here, each of the $\nu$ intervals contributes its diffusive resistance, but
this is diminished by the ballistic contributions of the $\nu - 1$ valleys,
which enter in the weight factors $ \exp[-2|x-(X -s/2)|/l]$ representing the
probabilities for collision-free motion.

The energy $E$ above which the tunneling probability $T^{\rm WKB}(E)$ becomes
appreciable is usually larger than $E_{\rm c}^{0}$.  The associated turning
points lie within a region of width $\Delta$, which defines the ``tunneling
interval'' \cite{lip01a}, i.e., the distance across which tunneling can occur
(cf.\ Fig.\ \ref{fig:1}).  Since tunneling is a ballistic process, one must
have $\Delta \ll l$, and we must exclude ballistic interval lengths smaller
than $\Delta$ when averaging over the configurations made up of these intervals
\cite{lip01a}.  This should entail a relative reduction of order $\Delta/l$ of
all integrals over $x$ in Eqs.\ (\ref{eq:5}), (\ref{eq:7}), and (\ref{eq:10}).
In the application considered in Sec.\ IV, we have $\Delta/l < 0.2$, and in
view of the other simplifications of the model we omit the corresponding
correction in our calculations.  This procedure may be viewed as a heuristic
description of the simultaneous occurrence of tunneling and diffusion.  It
differs from that of Ref.\ \cite{lip01a}, where the tunneling interval has been
excluded from the above-mentioned integrals.

\section{Electron mobilities}

In the unified transport model, the (effective) electron mobility $\mu$ for a
chain of $\nu$ identical grains of length $s$ is given  by
\begin{equation}
\mu = g \, \frac{\nu s}{e \bar{n}}  = \frac{4\pi e m^{*}}{\beta
h^{3}} \, \frac{\nu s}{\bar{n}} \, \ln (1+ e^{-\beta E_{\rm p}}) \,
\Gamma \, \frac{l}{L} \; ,
\label{eq:11}
\end{equation}
where $g$ is the conductance of Eq.\ (\ref{eq:1}) and $\bar{n}$ is the electron
concentration averaged over a grain.  The effective length $L$ depends
explicitly on the mean free path $l$ and on the number of grains, $\nu$ [cf.\
Eqs.\ (\ref{eq:4})--(\ref{eq:7})].  In particular,
\begin{equation}
\mu \propto \left\{
\begin{array}{ll}
\;  l/(l+\widetilde{s}) \;\;\; {\rm for} \;\; \nu = 1 \;\;\;\, ,  \\
\;  l/(\widetilde{\lambda}+\widetilde{s}) \;\;{\rm for} \;\; \nu \rightarrow
\infty \; .
\end{array}
\right.
\label{eq:11a}
\end{equation}
Note that the expression (\ref{eq:11}) refers to an ideal, one-dimensional
situation with strictly homogeneous grain boundaries.

Aside from the explicit dependence of the mobility $\mu$ on the mean free path
$l$, an implicit dependence on $l$ arises from the dependence of the band edge
profile $E_{\rm c}(x)$ on the donor concentration $N_{\rm d}$.  Adopting a
relation between $l$ and $N_{\rm d}$ obtained from Ref.\ \cite{aro82}, we
examine here the characteristic behavior of $\mu$ as a function of mean free
path $l$ (or donor concentration $N_{\rm d}$), grain number $\nu$, and grain
length $s$, and compare it to the behavior found for thermionic field emission.
For that mechanism, the mobility is obtained by setting $\nu = 1$ and $L=l$ in
Eq.\ (\ref{eq:11}) [the $l$-dependence of the mobility is then determined
solely by the implicit $l$-dependence of the band edge profile].

Using Fermi-Dirac statistics and assuming the grain boundaries to be
infinitesimally narrow, we calculate equilibrium band edge profiles $E_{\rm
c}(x)$ within the trapping model by numerically solving the nonlinear Poisson
equation (thereby going beyond the depletion approximation \cite{set75}) for a
single grain under periodic boundary conditions \cite{lip01a,wei99,lip01b}.
The density $D_{\rm t}(E)$ of trapping states at a grain boundary is taken as
the sum of a term corresponding to a deep level at energy $E_{0}$ and an
exponential describing tail states \cite{wer89} extending from the {\em bulk}
conduction band edge at $E_{\rm c}$ down into the gap, i.e.,
\begin{equation}
D_{\rm t}(E) = N_{0} \, \delta (E - E_{0}) + \kappa \, {\rm e}^{-(E_{\rm
c} - E)/\epsilon} \; .
\label{eq:12}
\end{equation}

\begin{figure}
\epsfysize=14.0cm
\epsfbox[32 450 280 646]{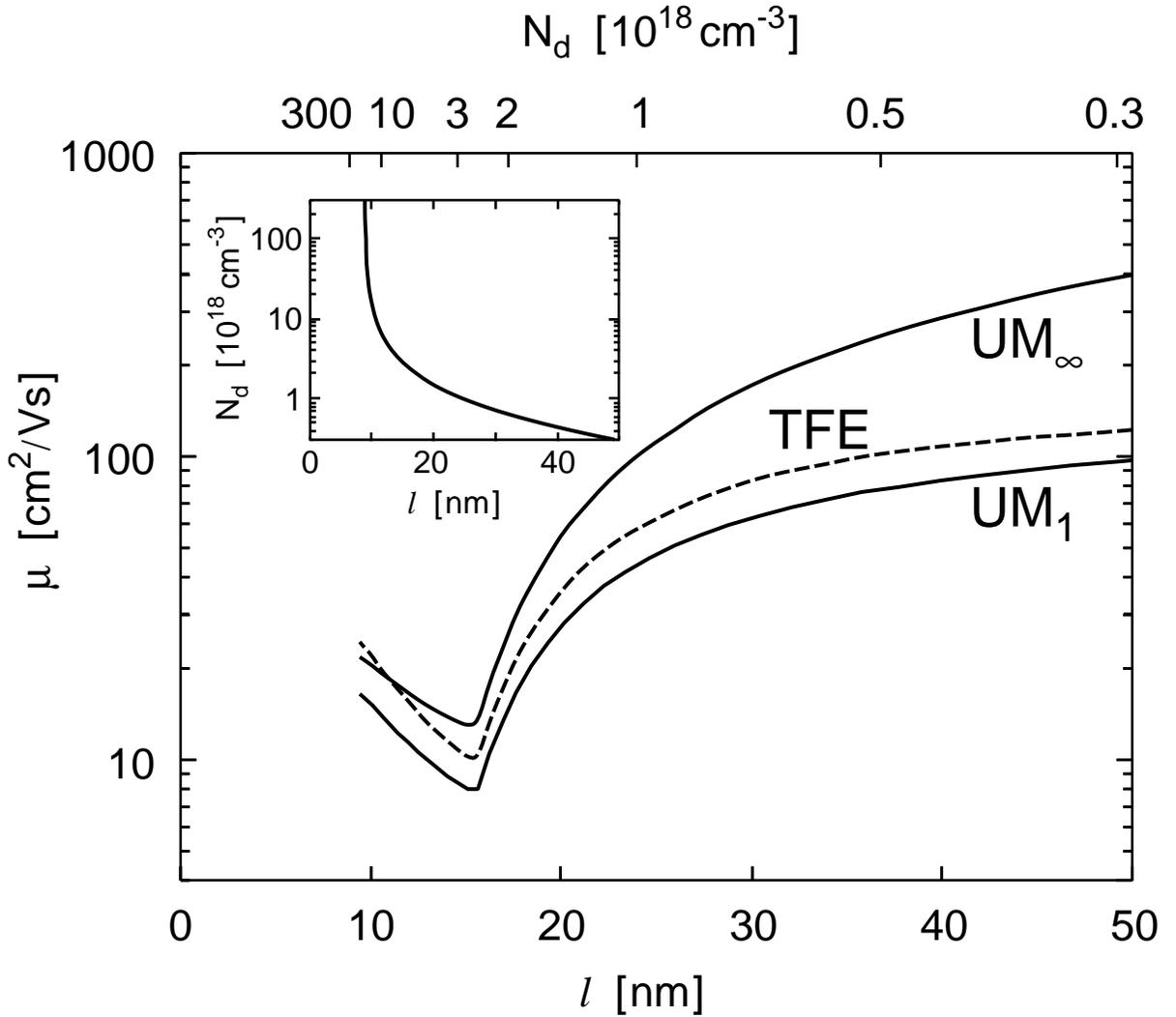}
\vspace*{1.7cm}
\caption{Theoretical electron mobilities for n-type $\mu$c-Si with grain
length $s = 15$ nm at temperature $T = 300$ K, plotted as a function of mean
free path $l$ (lower abscissa) [or of donor concentration $N_{\rm d}$ (upper
abscissa)].  In the conversion of $N_{\rm d}$ into $l$ (cf.\ inset), a relation
given in Ref.\ \protect\cite{aro82} was used.  Solid curves:\ mobilities
calculated from the unified-model expression (\ref{eq:11}); curve UM$_{1}$:\
single grain, curve UM$_{\infty}$:\ chain of many grains.  Dashed curve:\
mobility in the thermionic-field-emission (TFE) model.  For parameter values in
the trapping-state density (\ref{eq:12}), see text.
}
\label{fig:2}
\end{figure}

\begin{figure}
\epsfysize=14.0cm
\epsfbox[309 457 538 654]{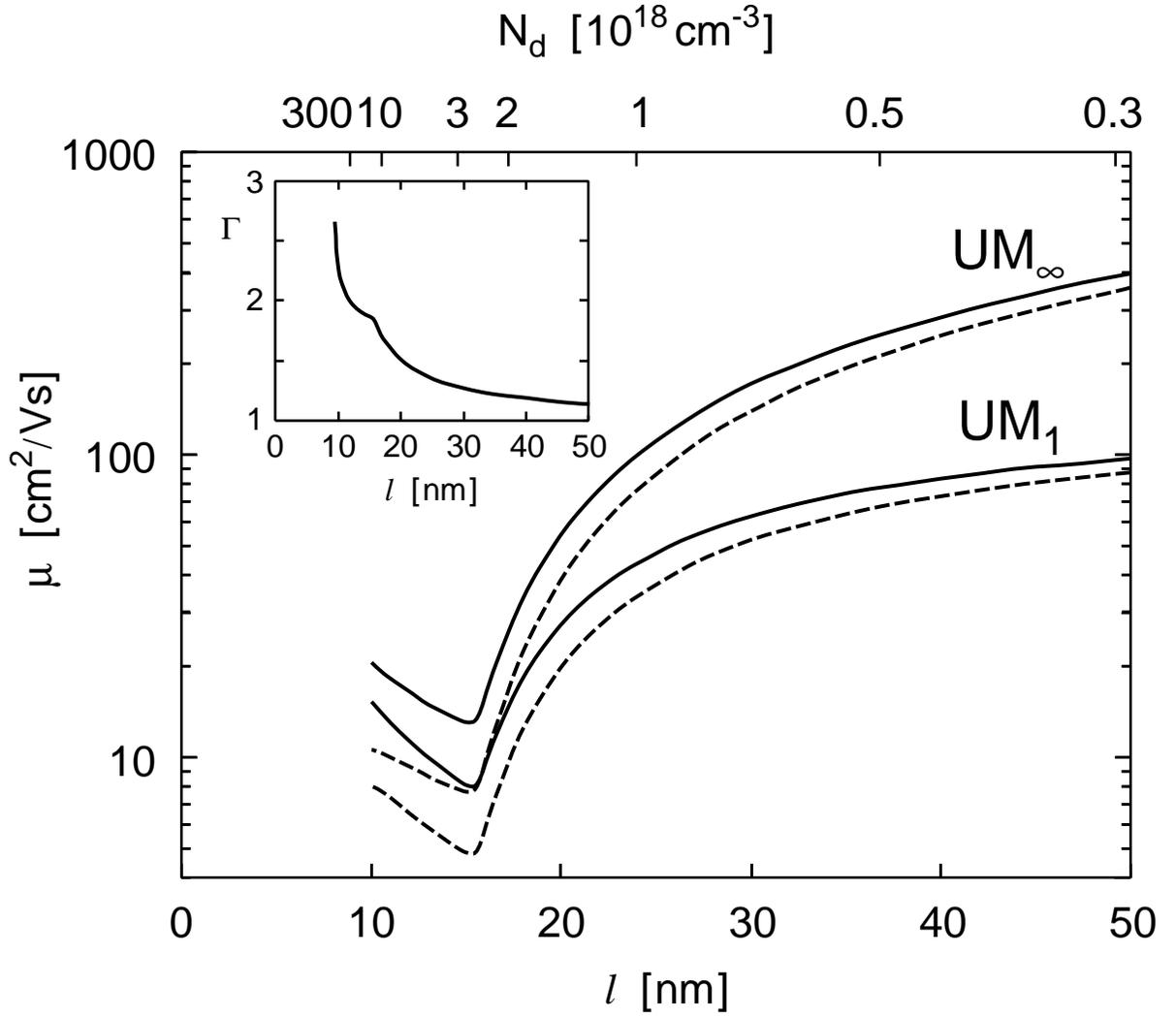}
\vspace*{1.8cm}
\caption{Illustration of the effect of tunneling on the electron mobilities.
Solid curves:\ unified-model mobilities of Fig.\ \ref{eq:2}.  Dashed curves:\
corresponding classical mobilities, calculated from Eq.\ (\ref{eq:11}) with
$\Gamma = 1$. Inset:\ $l$-dependence of the tunneling correction $\Gamma$.
}
\label{fig:2a}
\end{figure}

\begin{figure}
\epsfysize=11.0cm
\epsfbox[305 453 560 623]{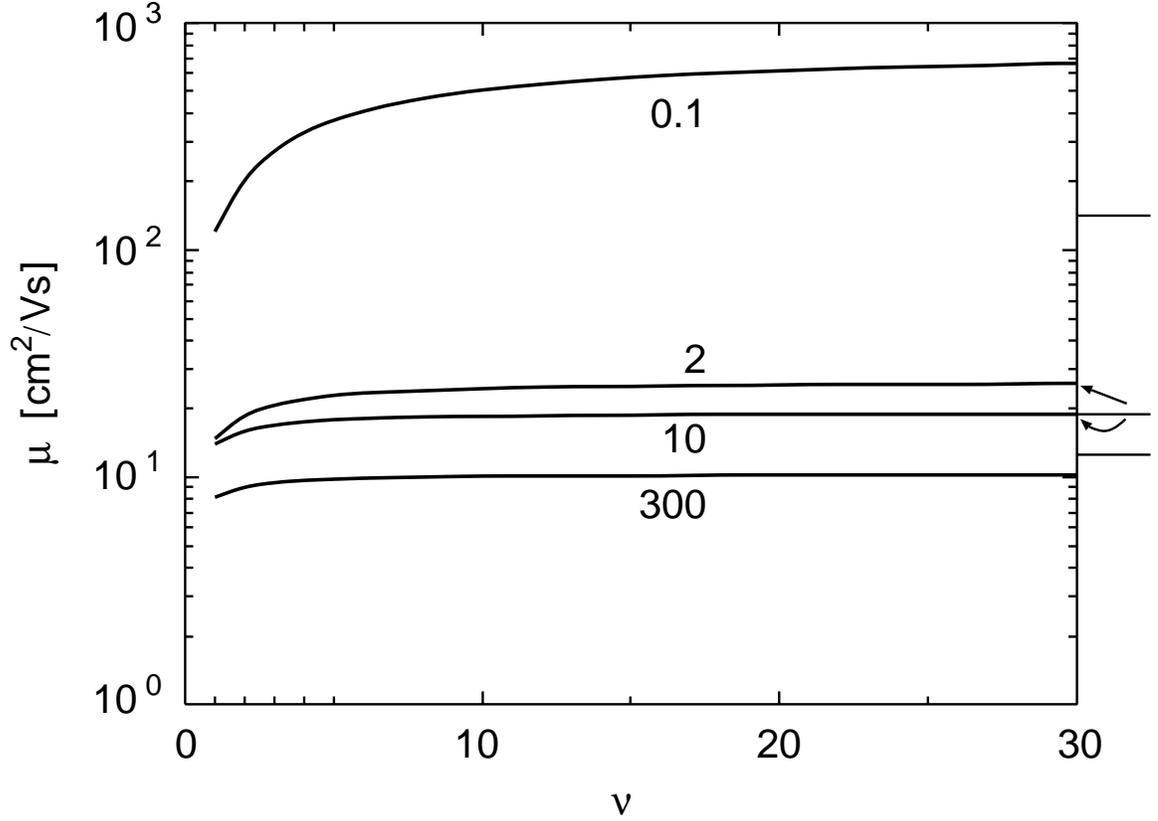}
\vspace*{1.7cm}
\caption{Unified-model mobilities for n-type $\mu$c-Si with grain length $s =
15$ nm at temperature $T = 300$ K, calculated from Eq.\ (\ref{eq:11}) as a
function of the number of grains, $\nu$, for different values of the donor
concentration $N_{\rm d}$.  The numbers attached to the curves are the $N_{\rm
d}$-values in units of 10$^{18}$ cm$^{-3}$.  For comparison, the corresponding
mobilities in the thermionic-field-emission model, with $N_{\rm d}$ increasing
from top to bottom, are indicated by bars at the right-hand ordinate.  For the
parameter values in the trapping-state density (\ref{eq:12}), see text.
}
\label{fig:3}
\end{figure}

\begin{figure}
\epsfysize=13.0cm
\epsfbox[37 219 259 393]{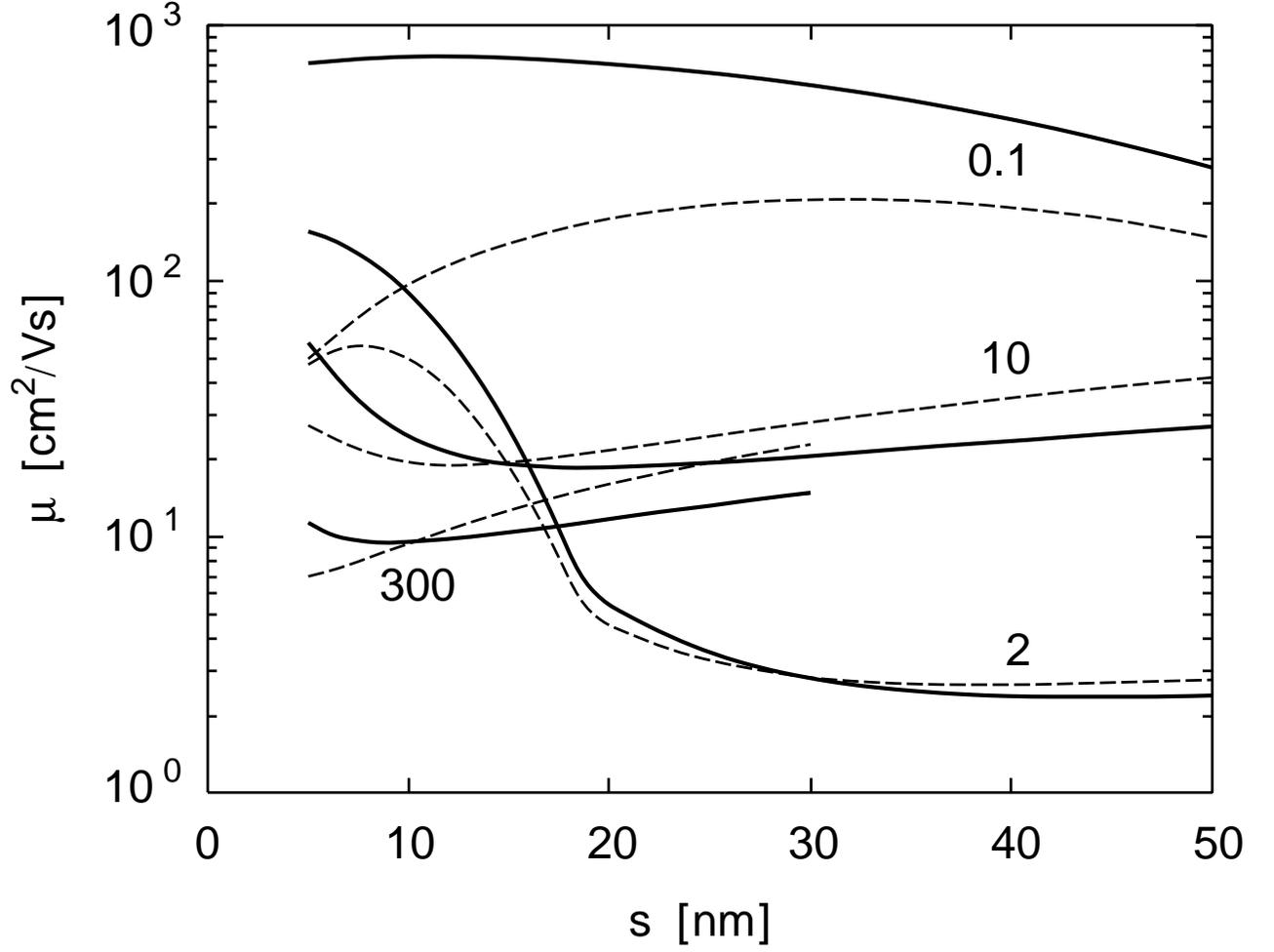}
\vspace*{1.75cm}
\caption{Theoretical electron mobilities for n-type $\mu$c-Si at temperature
$T = 300$ K and different donor concentrations $N_{\rm d}$, plotted as a
function of grain length $s$.  Solid curves:\ mobilities calculated from the
unified-model expression (\ref{eq:11}) for $\nu \rightarrow \infty$.  Dashed
curves:\ mobilities in the thermionic-field-emission model.  The numbers
attached to pairs of curves are the $N_{\rm d}$-values in units of 10$^{18}$
cm$^{-3}$.  For the parameter values in the trapping-state density
(\ref{eq:12}), see text.
}
\label{fig:4}
\end{figure}

Electron mobilities for n-type $\mu$c-Si at room temperature ($T=300$ K),
calculated from the unified-model expression (\ref{eq:11}), are displayed in
Figs.\ \ref{fig:2}--\ref{fig:4} as a function of the parameters $l$, $\nu$, and
$s$.  For the parameters in the trapping-state density $D_{\rm t}(E)$, we have
used values that emerged from our previous analysis of Hall mobility data for
phosphorus-doped $\mu$c-Si within the thermionic-field-emission model including
a fully quantal treatment of tunneling \cite{wei02}, viz., $N_{0} = \alpha_{0}
+ \alpha_{1} \, N_{\rm d}$ ($\alpha_{0} = 3.54 \times 10^{12}$ cm$^{-2}$,
$\alpha_{1} = 1.03 \times 10^{-7}$ cm), $E_{0} = (E_{\rm v} + E_{\rm c})/2$
($E_{\rm v}$ = bulk valence band edge)$, \kappa = 1 \times 10^{15}$ cm$^{-2}$
eV$^{-1}$, $\epsilon = 28$ meV.  For exhibiting characteristic features of the
unified model, this choice appears adequate.  We will reconsider the
determination of the parameter values in the trapping-state density in Sec.\ IV
where Hall mobility data are analyzed within the unified model.

In Fig.\ \ref{fig:2}, mobilities are shown as a function of mean free path $l$
(or of donor concentration $N_{\rm d}$) for a single grain ($\nu = 1$) and for
a chain of many grains ($\nu \rightarrow \infty$) [cf.\ Eq.\ (\ref{eq:11a})].
For comparison, the mobility of the thermionic-field-emission model is also
shown.  The pronounced minimum observed in all curves is associated with the
transition from complete depletion of the grains to partial depletion when the
donor concentration increases beyond a critical value \cite{set75}.  In the
range $l \approx s$, the unified-model mobility for $\nu \rightarrow \infty$ is
larger by about 50\%, on average, than that for $\nu = 1$.  For larger $l$, the
ratio of these two mobilities tends to diverge, with a variation proportional
to $l/s$ when $l \gg s$ [cf.\ Eqs.\ (\ref{eq:7}) and (\ref{eq:11a})].

The transport mechanism in the unified model becomes purely ballistic in the
large-$l$ limit.  Therefore, for a single grain, the corresponding mobility
shown in Fig.\ 2 approaches that for the thermionic-field-emission mechanism
which is also purely ballistic.  On the other hand, for a chain of grains, the
thermionic-field-emission model (and, of course, the thermionic-emission model
as well) identifies the conductivity, and hence also the mobility, of a single
grain with that of the whole chain.  It thus tacitly assumes that the mean free
path $l$ is long compared to the width of the barriers but short compared to
the length of the grains, so that while moving through the grains, the
electrons are thermalized and ``face'' each grain boundary barrier with the
same thermal distribution as when passing over the previous one (cf.\ Sec.\ 4.1
of Ref.\ \cite{lip01a}).  The thermionic-emission model is then no longer a
purely ballistic model; for $\nu > 1$, its results for large $l$ must differ
from those of the unified model, in which the barriers begin to ``eclipse''
each other when $l$ becomes larger than the distance $s$ between them.

The influence of tunneling on the unified-model mobilities is illustrated in
Fig.\ \ref{fig:2a}.  The tunneling correction $\Gamma$ given by Eq.\
(\ref{eq:2}) is seen to rise rapidly with decreasing mean free path $l$ in the
range $l < 30$ nm.  This behavior is traced to the rapid narrowing of the
grain-boundary barriers (and to the concomitant increase in their transparency)
when the donor concentration $N_{\rm d}$ increases beyond 10$^{18}$ cm$^{-3}$.
In Eq.\ (\ref{eq:11}), the quantity $\Gamma$ enters as an overall factor and
via the $\Gamma$-dependence of the effective length $L$ in the denominator
[cf.\ Eqs.\ (\ref{eq:4}), (\ref{eq:5}), and (\ref{eq:7})].  It appears that the
latter dependence has small influence only, and thus tunneling enhances the
mobility essentially by a factor $\Gamma$ over its classical value.

In Fig.\ \ref{fig:3}, unified-model mobilities are shown as a function of the
number of grains, $\nu$.  Sizeable variations of the mobility are restricted to
the range of small $\nu$ and very low donor concentration $N_{\rm d}$.  The
$\nu$-dependence of the mobility arises from the eclipsing effect, which is
most notable when comparing the transport across one barrier ($\nu=1$) with
that across two or three barriers ($\nu = 2,3$).  As the number of barriers
increases, the eclipsing effect of the added barriers becomes negligible, and
the mobility becomes independent of $\nu$ [cf.\ Eq.\ (\ref{eq:11a})] (but {\em
not} equal to that for a single grain).

The dependence of the mobilities of the unified and the
thermionic-field-emission models on the grain length $s$ is displayed in Fig.\
\ref{fig:4}.  A rapid decrease of the mobility with increasing $s$ is observed
at low donor concentration close to the minimum in the $N_{\rm d}$-dependence
of $\mu$ appearing in Fig.\ \ref{fig:2} (cf.\ also Ref.\ \cite{set75}).  For
very low $N_{\rm d}$, there are strong discrepancies between the results of the
unified model and those of the thermionic-field-emission model, in particular
in the range of small~$s$.

For large numbers of grains ($\nu \rightarrow \infty$) and for mean free paths
comparable to the grain length $(l \approx s$), i.e., for parameter values
encountered in many experimental studies, the results of Figs.\
\ref{fig:2}--\ref{fig:4} show that (i) the unified-model mobility is fairly
close to the mobility of the thermionic-field-emission model; (ii) tunneling
accounts for up to 50\% of the unified-model mobility; (iii) the unified-model
mobility is virtually independent of the number of grains; (iv) as a function
of grain length, the unified-model mobility exhibits rapid variations when the
donor concentration is low.

\section{Application: Analysis of Hall mobilities}

In this section, we apply the unified transport model in an analysis of Hall
mobility data for thin films of n-type $\mu$c-Si.  Following common practice
\cite{ort80a}, we identify the effective mobility $\mu$ given by expression
(\ref{eq:11}) with the Hall mobility $\mu_{\rm H}$.  By fitting, in the range
of thermally activated transport, the temperature dependence of $\mu$ to that
of $\mu_{\rm H}$, we infer {\em true} potential barrier heights, which solely
reflect structural properties of the grain boundaries.  The deviations of the
Hall activation energies from the true barrier heights exhibit the effect of
the transport mechanism on the mobilities.

\subsection{Activation energies and barrier heights}

Our analysis is based on Hall data obtained for samples of highly
phosphorus-doped $\mu$c-Si:H thin films, with thickness varying between 0.15
and 0.28 $\mu$m and average grain length $s \approx 15$ nm (for details, see
Refs.\ \cite{bre01a,bre01b}).  In Fig.\ \ref{fig:5}, the temperature dependence
of the Hall mobilities $\mu_{\rm H}$ is shown in an Arrhenius plot.  Fitting
straight lines to the mobilities in the high-$T$ range, i.e., assuming the
mobilities there to have the form
\begin{equation}
\mu_{\rm H} = \mu_{\rm H}^{0} \, \exp(-E_{\rm a}/k_{\rm B}T) \; ,
\label{eq:13}
\end{equation}
we have determined values for the preexponentials $\mu_{\rm H}^{0}$ and the
activation energies $E_{\rm a}$ for the different samples (by restricting
ourselves to the high-$T$ range, we avoid entering a discussion of the
intricacies of curved Arrhenius behavior; cf.\ Ref.\ \cite{wer94}, and
references cited therein).  In Table I, the values of $\mu_{\rm H}^{0}$ and
$E_{\rm a}$ are listed along with donor concentrations $N_{\rm d}$ estimated
\cite{bre01a} from the electron concentrations by using Seto's model
\cite{set75}.  The mean free paths $l$ corresponding to the $N_{\rm d}$-values
vary between 10 nm and 20 nm (cf.\ inset in Fig.\ \ref{fig:2}), i.e., the
condition $l \approx s$ is fulfilled.

\begin{figure}
\epsfysize=12.5cm
\epsfbox[318 215 548 390]{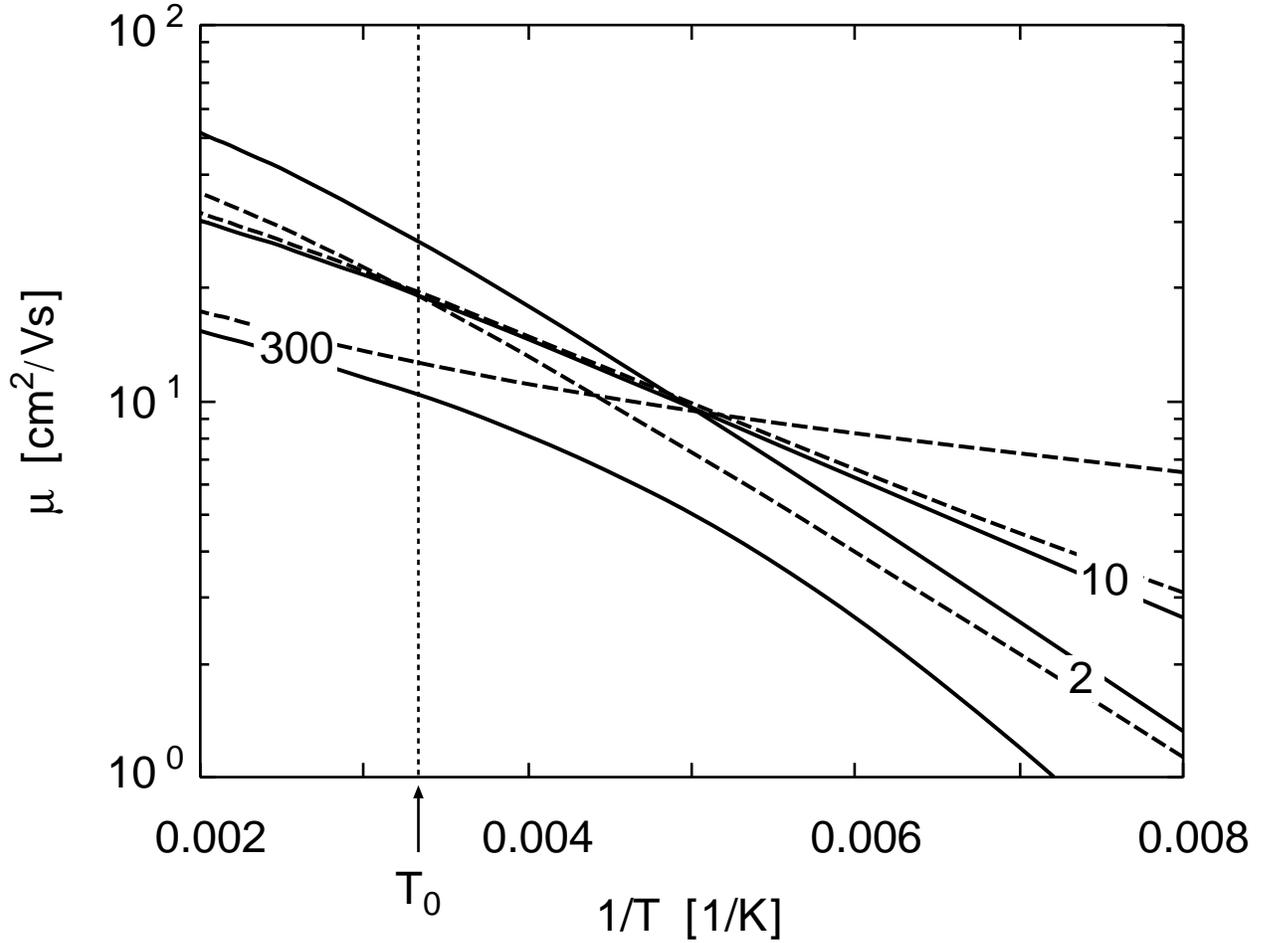}
\vspace*{1.7cm}
\caption{Theoretical electron mobilities for n-type $\mu$c-Si with grain
length $s = 15$ nm and different donor concentrations $N_{\rm d}$, plotted as a
function of inverse temperature, $1/T$.  Solid curves:\ mobilities calculated
from the unified-model expression (\ref{eq:11}) for $\nu \rightarrow \infty$.
Dashed curves:\ mobilities in the thermionic-field-emission model.  The numbers
attached to pairs of curves are the $N_{\rm d}$-values in units of 10$^{18}$
cm$^{-3}$.  For the parameter values in the trapping-state density
(\ref{eq:12}) and for the definition of $T_{0}$, see text.
}
\label{fig:6}
\end{figure}

In Fig.\ \ref{fig:6}, the temperature dependence of the mobilities of the
unified model (for grain number $\nu \rightarrow \infty$) and the
thermionic-field-emission model for grain length $s = 15$ nm and different
values of the donor concentration $N_{\rm d}$ is displayed in an Arrhenius
plot. For the parameters in the trapping-state density $D_{\rm t}(E)$, the
same values as in the calculations shown in Figs.\ \ref{fig:2}--\ref{fig:4}
were used.  Major deviations from pure Arrhenius behavior are observed only for
the unified-model results at $N_{\rm d} = 3 \times 10^{20}$ cm$^{-3}$ and low
temperatures.  These may be attributed to the strong temperature dependence of
the Fermi level in that parameter range and its effect on the terms
$\widetilde{\lambda}$ and $\widetilde{s}$ [cf.\ Eqs.\ (\ref{eq:11}) and
(\ref{eq:11a})].  In those temperature ranges where the different data sets in
Fig.\ \ref{fig:5} exhibit Arrhenius behavior, the temperature dependence of the
corresponding theoretical mobilities can be well approximated, in the vicinity
of a fixed temperature $T_{0}$, by an Arrhenius formula
\begin{equation}
\mu = \mu_{0} \, \exp(-E_{\rm b}^{\rm eff}/k_{\rm B}T) \; ,
\label{eq:14}
\end{equation}
where the preexponential $\mu_{0}$ and the ``effective barrier height'' $E_{\rm
b}^{\rm eff}$ depend weakly on $T_{0}$.  We then can identify $E_{\rm b}^{\rm
eff}$, evaluated at a suitably chosen $T_{0}$-value, with the activation energy
$E_{\rm a}$ derived from the experimental data.  In the following, we use
$T_{0} = 300$ K throughout; this value is roughly equal to the average of the
centers of those $T$-intervals over which the Hall data of Fig.\ \ref{fig:5}
exhibit Arrhenius behavior.

In order to infer true barrier heights, we match the effective barrier heights
$E_{\rm b}^{\rm eff}$ to the activation energies $E_{\rm a}$ given in Table I
by adjusting the values of the parameters in the trapping-state density $D_{\rm
t}(E)$.  In our previous analysis of Hall mobilities within the
thermionic-field-emission model \cite{wei02}, we have assigned fixed values to
the parameters $E_{0}$, $\kappa$, and $\epsilon$, and subsequently adjusted the
value of the area density $N_{0}$ independently for each sample, i.e., for each
donor concentration $N_{\rm d}$, such as to obtain $E_{\rm b}^{\rm eff} =
E_{\rm a}$.  The resulting $N_{0}$-values can be accurately fitted by means
of the interpolation formula $N_{0} = \alpha_{0} + \alpha_{1} \, N_{\rm d}$
(cf.\ Sec.\ III).

Here, within the unified model, we adopt the matching procedure of Ref.\
\cite{wei02}, using, in particular, the same values for the parameters $E_{0}$,
$\kappa$, and $\epsilon$ (cf.\ Sec.\ III).  The adjusted values of the area
density $N_{\rm 0}$ for the different donor concentrations $N_{\rm d}$ can
again be interpolated by a linear expression, with parameter values $\alpha_{0}
= 3.79 \times 10^{12}$ cm$^{-2}$, $\alpha_{1} = 7.04 \times 10^{-8}$ cm.  While
the value of $\alpha_{0}$ is close to that of Ref.\ \cite{wei02}, the value of
$\alpha_{1}$ is smaller by about 30\%.  Varying the values of $E_{0}$,
$\kappa$, and $\epsilon$ within reasonable bounds does not lead to a
substantial change in the adjusted values of $N_{0}$.

\begin{figure}
\epsfysize=12.5cm
\epsfbox[55 222 283 395]{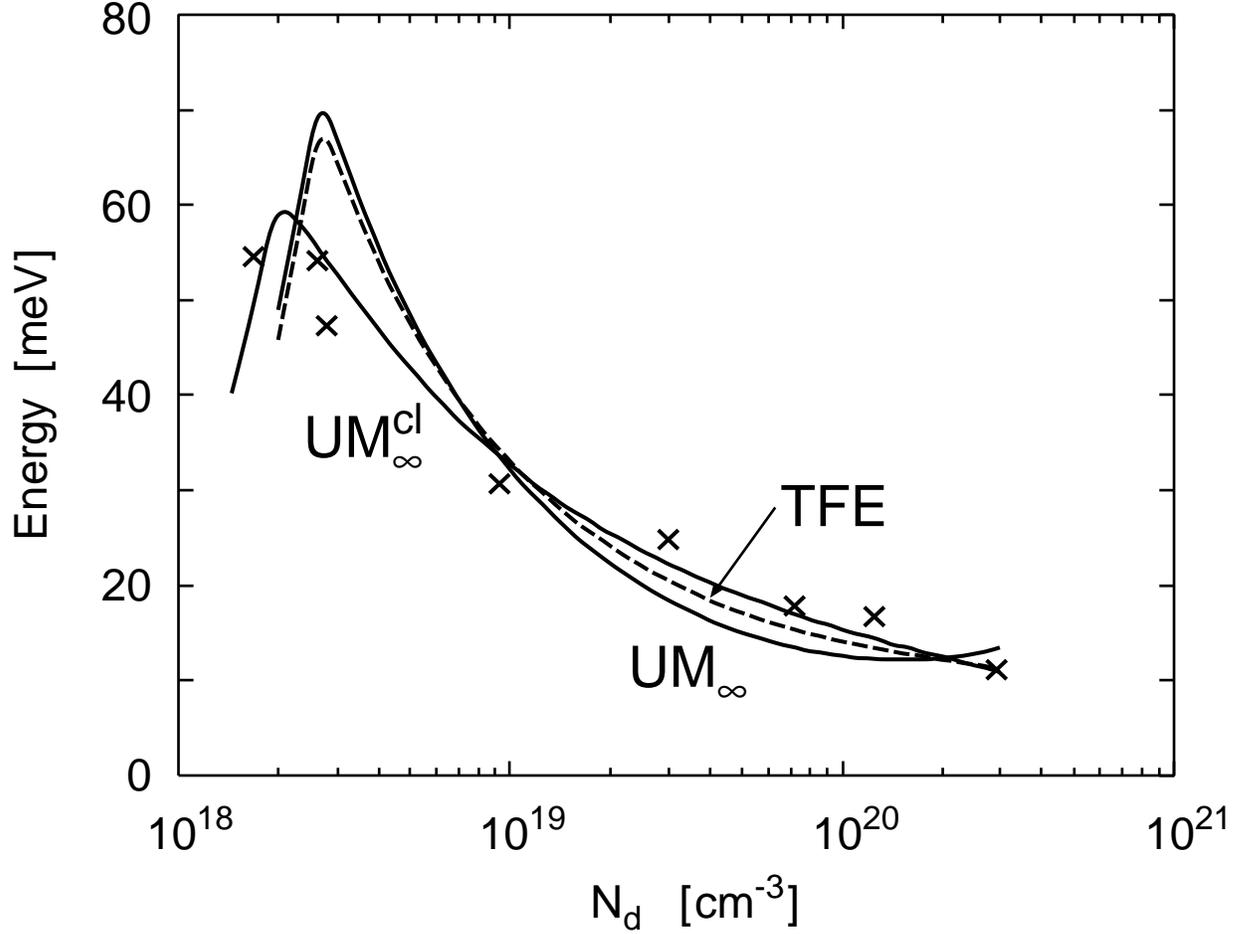}
\vspace*{1.7cm}
\caption{Activation energies and effective barrier heights for n-type
$\mu$c-Si as a function of donor concentration $N_{\rm d}$.  Symbols:\
activation energies $E_{\rm a}$ deduced from the Hall data (cf.\ Fig.\
\ref{fig:6} and Table I).  Solid curves:\ effective barrier heights $E_{\rm
b}^{\rm eff}$ for $s = 15$ nm, obtained by fitting expression (\ref{eq:14}) to
the unified-model mobilities calculated from Eq.\ (\ref{eq:11}) at $T_{0}=300$
K; curve UM$_{\infty}$:\ result based on the full expression (\ref{eq:11}),
curve UM$_{\infty}^{\rm cl}$:\ classical result [$\Gamma = 1$ in Eq.\
(\ref{eq:11})].  Dashed curve:\ effective barrier heights obtained from the
thermionic-field-emission model.  For the parameter values in the
trapping-state density (\ref{eq:12}), see text.
}
\label{fig:7}
\end{figure}

Using the interpolation formula for $N_{0}$ as a function of $N_{\rm d}$, we
have calculated from Eqs.\ (\ref{eq:11}) and (\ref{eq:14}) a continuous curve
for the $N_{\rm d}$-dependence of the effective barrier height $E_{\rm b}^{\rm
eff}$, which is displayed in Fig.\ \ref{fig:7} along with the activation
energies $E_{\rm a}$.  Also shown are the results of the
thermionic-field-emission model, calculated with the same set of parameter
values as used in the unified-model calculations.  In obtaining the {\em
classical} unified-model results for $E_{\rm b}^{\rm eff}$ [$\Gamma = 1$ in
Eq.\ (\ref{eq:11})] included in Fig.\ \ref{fig:7}, an independent adjustment of
the parameter values in the trapping-state density $D_{\rm t}$ was performed.
With $\kappa = 5 \times 10^{14}$ cm$^{-2}$ eV$^{-1}$ and $\epsilon = 30$ meV, a
good fit to the activation energies was achieved by choosing for the area
density $N_{0}$ the constant ($N_{\rm d}$-independent) value $2.89 \times
10^{12}$ cm$^{-2}$.

The true grain-boundary barrier height $E_{\rm b}$, defined here as $E_{\rm b}
= E_{\rm c}^{\rm m} - E_{\rm c}^{0}$ (i.e., identified with the ``band
bending'' \cite{ort80a}), is directly read from the band edge profile $E_{\rm c
}(x)$ calculated with the adjusted parameter values in the trapping-state
density.  Referring the barrier height to the profile at the center of the
grain, $E_{\rm c}^{0}$, is appropriate when discussing results in which
tunneling is taken into account [cf.\ Eq.\ (\ref{eq:2})].  In Fig.\
\ref{fig:8}, we show barrier heights $E_{\rm b}$ for the full and the classical
unified model as a function of donor concentration $N_{\rm d}$.  For direct
comparison, we include the corresponding results for $E_{\rm b}^{\rm eff}$
(cf.\ Fig.\ \ref{fig:7}).

\subsection{Discussion}

\begin{figure}
\epsfysize=12.5cm
\epsfbox[316 222 542 395]{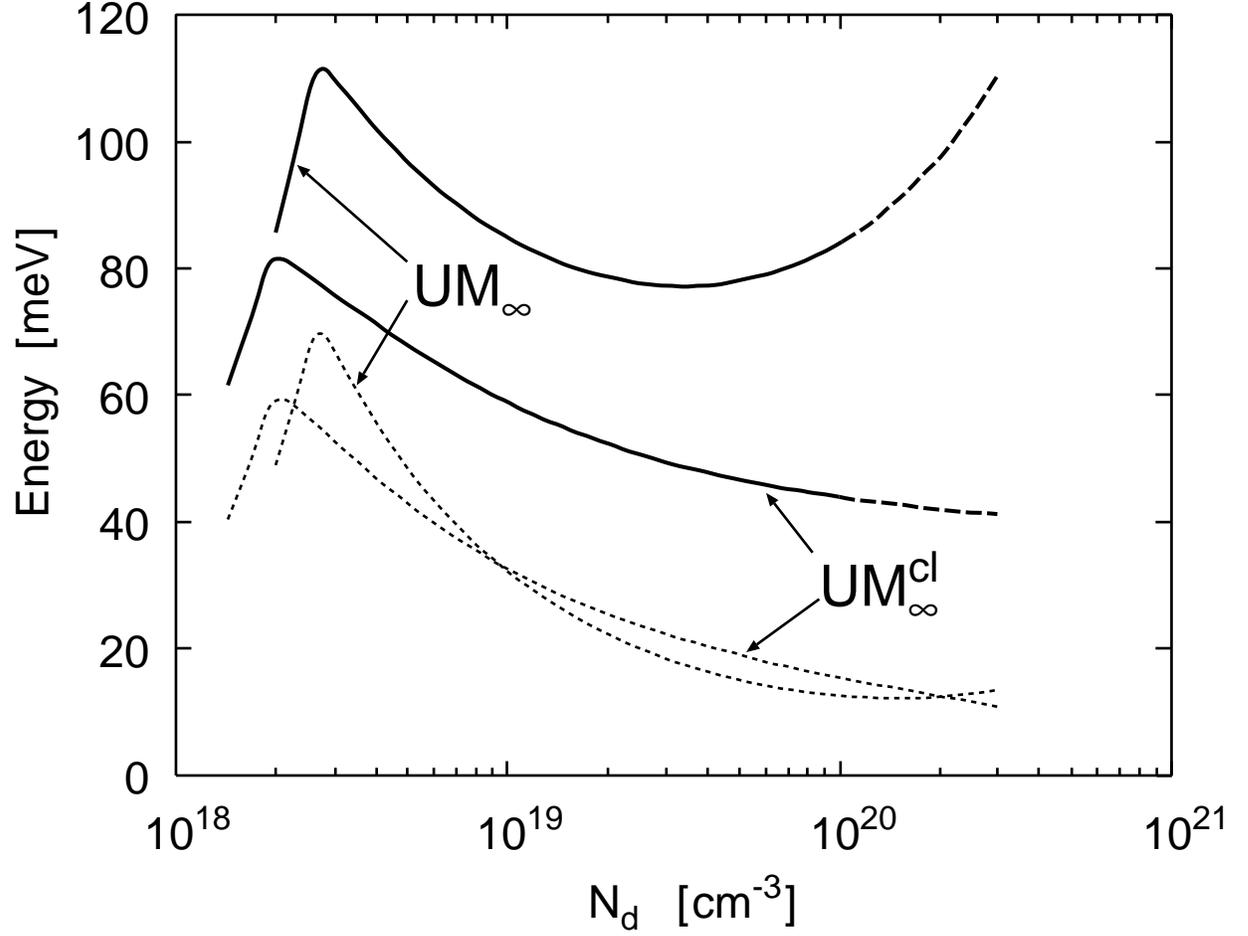}
\vspace*{1.6cm}
\caption{Grain-boundary barrier heights for n-type $\mu$c-Si as a function of
donor concentration $N_{\rm d}$.  Solid curves:\ true barrier heights $E_{\rm
b}$ for the parameter values of Fig.\ \ref{fig:7}, derived within the unified
transport model; curve UM$_{\infty}$:\ result based on the full expression
(\ref{eq:11}), curve UM$_{\infty}^{\rm cl}$:\ classical result [$\Gamma = 1$ in
Eq.\ (\ref{eq:11})]; the meaning of the dashed parts of the curves is explained
in the text.  Dotted curves:\ corresponding effective barrier heights $E_{\rm
b}^{\rm eff}$ (cf.\ Fig.\ \ref{fig:7}).
}
\label{fig:8}
\end{figure}

The principal result of our analysis of Hall mobilities is that the true
potential barrier heights derived within the unified model deviate
substantially from the Hall activation energies.  Considering the results
presented in Figs.\ \ref{fig:7} and \ref{fig:8} in detail, we conclude that
although the unified and thermionic-field-emission models generally yield
appreciably different values for the electron mobilities (cf.\ Sec.\ III), the
effective barrier heights obtained from the two models are in close agreement
when one and the same (adjusted) set of parameters in the trapping-state
density is used, thus implying identical values for the {\em true} barrier
heights.  In the present analysis, therefore, the effective barrier heights
turn out to be not sufficiently sensitive to either type of transport
mechanism; the subtle effects of the interplay between ballistic and diffusive
transport, which are taken into account in the unified model, cannot be
resolved within the uncertainties of the analysis.  In the unified model, as in
the thermionic-field-emission model, the transport mechanism manifests itself
essentially through the effect of tunneling, which causes the true barrier
heights to deviate appreciably from the effective heights, in particular for
large donor concentrations.

It is seen that the true barrier heights derived within the classical
approximation come closer to the effective heights, i.e., to the Hall
activation energies.  Finally, it is found that the adjusted values of the area
density of the trapping states rise rapidly with increasing donor concentration
when tunneling is taken into account.  In the classical description, on the
other hand, the area density exhibits a very weak dependence on donor
concentration.  The latter result is in line with the finding obtained by Seto
\cite{set75} for p-type $\mu$c-Si thin films.

In order to assess the general reliability of the present description and the
relevance of the results of the analysis of Hall data, a number of remarks are
in order.

(i) In our description of barrier-controlled transport, we have disregarded a
number of effects whose consideration may quantitatively alter our results.  We
have confined ourselves to a one-dimensional formulation and have treated
chains of identical grains only.  Fluctuations in grain length, the
non-planarity and inhomogeneity of the grain boundaries, and the perturbation
caused by column boundaries have not been taken into account.

(ii) The absolute values of the electron mobilities calculated for n-type
$\mu$c-Si are by about one order of magnitude larger than the experimental
values (cf.\ Figs.\ \ref{fig:5} and \ref{fig:6}).  Inhomogeneities in the
grain-boundary properties that are not allowed for in our one-dimensional
treatment are likely to be responsible \cite{kam86,pri98} for this
feature.

(iii) There is a zone of disorder of finite extension at real grain boundaries
(albeit estimated to be very narrow (width $< 1$ nm) in $\mu$c-Si
\cite{gro85}).  Thus, the band edge profiles calculated under the assumption of
infinitesimally narrow grain boundaries will have no physical meaning in the
close vicinity of the boundary.  Accordingly, with increasing narrowing of
the barriers, i.e., with increasing donor concentration, the uncertainty in the
values inferred for the true barrier height $E_{\rm b}$ rises.  For $N_{\rm d}
> 10^{20}$ cm$^{-3}$, the calculated widths of the barriers are smaller than 1
nm, so that in this range no significance can be attached to the values of
$E_{\rm b}$ (cf.\ Fig.\ \ref{fig:8}, dashed parts of upper curves).

(iv) In our analysis of Hall mobilities, uncertainties (aside from those
associated with the basic limitations of our theoretical treatment) arise
mainly from inaccuracies in the $N_{\rm d}$-values assigned to the different
samples.  The finite film thickness and its variation from sample to sample may
also be a source of uncertainty.

\section{Summary and conclusions}

In this paper, carrier transport across grain-boundary potential barriers in
microcrystalline semiconducting materials has been theoretically studied with
the aim to elucidate the role played by the transport mechanism.  To this end,
a recently developed model that unifies the ballistic and diffusive mechanisms
was applied in the calculation of electron mobilities.  A one-dimensional
formulation of the model was used in conjunction with the trapping model.
Quantum corrections allowing for tunneling were taken into account within the
WKB approximation.

In order to exhibit general features of the unified model, electron mobilities
for $\mu$c-Si have been calculated as a function of various parameters, such as
carrier mean free path, grain length, number of grains, and temperature.
Sizeable deviations from the results of the thermionic-field-emission model
were found, particularly for large numbers of grains and for small grain
lengths at low donor concentration.

Furthermore, we have confronted the unified transport model with experimental
results by analyzing Hall mobilities for highly phosphorus-doped $\mu$c-Si:H
thin films in the range of thermally activated transport.  From the Hall
activation energies, ``true'' grain-boundary barrier heights were determined by
matching effective barrier heights deduced from the unified-model mobilities to
the activation energies.  The true heights turn out to be substantially larger
than the activation energies, mainly owing to the effect of tunneling.  The
area density of trapping states obtained from the matching procedure rises
rapidly with increasing donor concentration if tunneling is included in the
unified-model calculation, but is virtually independent of donor concentration
in the classical case.

In view of the limitations of our model description of carrier transport and of
the uncertainties of our analysis of Hall mobilities, it would be premature to
draw general conclusions from the results of the present study.  Our results
indicate, however, that in many cases of practical interest the
thermionic-field-emission model may provide a sufficiently good description of
barrier-controlled carrier transport.  In order to reach definitive conclusions
on the relevance of the unified transport model, more general versions of this
model have to be worked out and systematically applied in the analysis of
transport properties of microcrystalline semiconducting materials.

\begin{figure}
\epsfysize=13.0cm
\epsfbox[56 454 281 630]{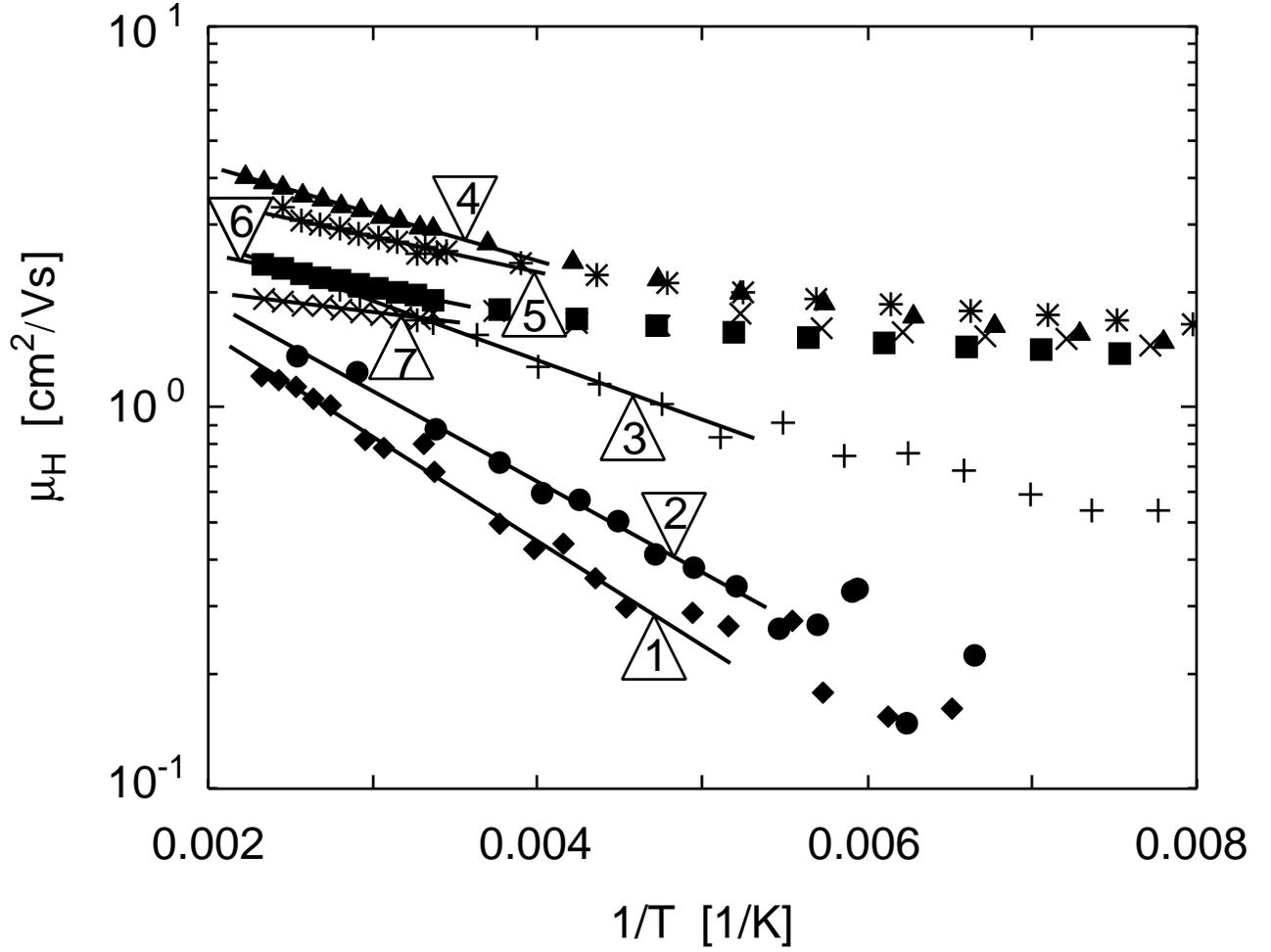}
\vspace*{1.8cm}
\caption{Hall mobilities as a function of inverse temperature for thin films of
phosphorus-doped $\mu$c-Si:H \protect\cite{bre01a,bre01b}.  The straight lines
fitted to the data in the high-$T$ range determine the preexponentials
$\mu_{\rm H}^{0}$ and the activation energies $E_{\rm a}$ introduced in Eq.\
(\ref{eq:13}).  The labels attached to the different lines correspond to the
numbering of the samples in Table I.
}
\label{fig:5}
\end{figure}

\newpage
\begin{center}
{\large TABLES}\\[1.0cm]
\end{center}
TABLE I.\ Preexponentials $\mu_{\rm H}^{0}$ and activation energies $E_{\rm a}$
for different samples of phosphorus-doped $\mu$c-Si:H thin films with donor
concentrations $N_{\rm d}$, obtained by fitting expression (\ref{eq:13}) to the
Hall mobility data shown in Fig.\ \ref{fig:5}.\\[1.0cm]
\begin{center}
\begin{tabular}{|c||c|c|c|} \hline
\# & $N_{\rm d}$ (10$^{18}$ cm$^{-3}$)  & $\mu_{\rm H}^{0}$ (cm$^{2}$ V$^{-1}$
s$^{-1}$)  & $E_{\rm a}$ (meV) \\ \hline
1 & 1.7 & 5.6 &  55  \\
2 & 2.8 & 5.6 &  47  \\
3 & 9.3 & 5.5 &  31  \\
4 & 30  & 7.7 &  25  \\
5 & 72  & 5.2 &  18  \\
6 & 125 & 3.7 &  17  \\
7 & 292 & 2.6 &  11  \\ \hline
\end{tabular}
\end{center}

\newpage

\newpage

\newpage

\newpage

\newpage

\newpage

\newpage

\newpage

\end{document}